\documentclass[12pt]{iopart}
\usepackage{iopams}  
\usepackage{amssymb}
\usepackage{setstack}  
\usepackage{graphicx}
\usepackage{bm}
\usepackage{amsbsy}
\usepackage{amstext}
\usepackage[T1]{fontenc}
\usepackage[latin9]{inputenc}

\newcommand{\Op}[1]{{\boldsymbol{\mathrm{\hat{#1}}}}}

\begin{document}

\title[Exceptional points for parameter estimation: The Bloch equations]
{Exceptional points for parameter estimation in open quantum systems: 
	Analysis of the Bloch equations}

\author{Morag  Am-Shallem,  Ronnie Kosloff}

\ead{ronnie@fh.huji.ac.il}

\address{Fritz Haber Research Center and the Institute of Chemistry,
	the Hebrew University, Jerusalem 91904, Israel}
\author{Nimrod Moiseyev}
\address{Schulich Faculty of Chemistry and Faculty of Physics, Technion, Haifa 3200008, Israel}

\begin{abstract}
We suggest to employ the dissipative nature of open quantum systems for the purpose of parameter estimation:
The dynamics of open quantum systems is typically described by a quantum dynamical semigroup generator ${\cal L}$.
The eigenvalues of ${\cal L}$ are complex, reflecting unitary as well as dissipative dynamics.  
For certain values of parameters defining ${\cal L}$,  non-hermitian degeneracies emerge, i.e. exceptional points ($EP$).
The dynamical signature of these {\em EP}s corresponds to a unique time evolution. 
This unique feature can be employed experimentally to locate
the {\em EP}s  and thereby to determine  the intrinsic system parameters  with a high accuracy.
This way we turn the disadvantage of the dissipation into an advantage. 
We demonstrate this method in the open system dynamics of a two-level system described by the Bloch equation,
which has become the paradigm of diverse fields in physics, from NMR to quantum information and elementary particles.
\end{abstract}

\section{Introduction}
\label{sec:intro}

Felix Bloch \cite{bloch1946nuclear} pioneered the dynamical description of open quantum systems.
Originally Bloch's equations describe the relaxation and dephasing of a nuclear spin in a magnetic field.
Soon it became apparent that the treatment can be extended to a generic two-level-system (TLS),
such as the dynamics of laser driven atoms in the optical regime \cite{agarwal1970master,cohen2008optical,Brewer1983bloch}.
The open TLS has been used to model many different fields of physics.
The TLS or a q-bit  is  at the foundation of quantum information \cite{lloyd1995almost,zrenner2002coherent,gammelmark2014fisher,clarke2008superconducting,ladd2010quantum}.
In particle physics the TLS algebra has been employed in
studies of possible deviations from quantum mechanics in the context of
neutrino oscillations \cite{lisi2000probing}, as well as
quantum entanglement \cite{six82,Selleri83,privitera1992quantum,datta1986quantum,PhysRevB.90.054304},
associated with electron/positron collisions and entangled
systems due to EPR-Bell correlations \cite{bell2004speakable}. 

The TLS is the base for setting the frequency standard for atomic clocks \cite{essen1955atomic}. 
As a result accurate measurement of frequency is an important issue. 
Quantum-enhanced measurements based on interferometry have been suggested as means to beat the shot noise limit \cite{giovannetti2004quantum}. 
In these methods the decoherence rate is the limiting factor \cite{cirac1997improve}. 
In some cases quantum error correction can increase the coherence time and the accuracy \cite{retzker2014increase}. 
In the present study we want to suggest an opposite strategy. 
By employing the non-hermitian character of the dynamics, 
the decoherence can be transformed from a bug  to a feature.

\section{Exceptional points in open quantum systems}
\label{sec:EP_in_LGKS}
The Bloch equation is the simplest example of a quantum Master equation. 
Bloch  rederived the equation from first principles, 
employing the assumption of weak coupling between the system and bath 
\cite{wangsness1953dynamical,bloch1957generalized}.
These studies have paved the way for a general theory of quantum open systems. 
Davies \cite{davies1974markovian} rigorously derived the weak coupling limit, 
resulting in a quantum Master equation which leads to a completely positive dynamical semigroup 
\cite{kraus1983states}.
Based on a mathematical construction, Lindblad and Gorini, Kossakowski and Sudarshan (L-GKS) obtained the general structure of the generator ${\cal L}$ of a completely positive dynamical semigroup \cite{lindblad1976generators,gorini1976completely}.
In the Heisenberg representation the L-GKS generator becomes \cite[Chapter 3]{breuer2002theory}:
\begin{equation}
\frac{d}{dt}\Op X~= 
~\frac{\partial \Op X }{\partial t}
+ i\left[\Op{H},\Op X \right]
+ \sum_{k}\left(
\Op{V}_{k}^{\dagger}\Op X \Op{V}_{k}
- \frac{1}{2} \left[ \Op{V}_{k}^{\dagger}\Op{V}_{k},\Op X \right] _+ 
\right).
\label{eq:LGKS}
\end{equation}
where $\Op X$ is an arbitrary operator.
The hamiltonian $\Op H$ is hermitian  and operators $\Op V_k$ are defined to operate in the Hilbert space of the system.
The $ \left[ \cdot, \cdot \right] _{+} $ denotes an anti commutator.

The set of operators $\lbrace \Op X \rbrace$ supports a Hilbert space construction 
using the scalar product:
$( \Op{X}_1 , \Op{X}_2) \equiv 
tr \lbrace \Op{X_1}^\dagger \Op{X_2}\rbrace
$.
A crucial simplification to Eq. (\ref{eq:LGKS}) is obtained when
a set of operator is closed to the generator $\cal{L}$.
Then we can rephrase the dynamics with a matrix-vector notation \cite{mukamel1999principles}:

\begin{equation}
%\frac{d}{dt} \vec Y =  M \vec Y
\dot{\vec Y} =  M \vec Y
\end{equation}
where $\vec Y$ is the vector of basis operators and $ M$ is the representation of the generator ${\cal L}$ in this vector space.
The eigenvalues of the matrix $ M$ reflect the non-hermitian dynamics generated by ${\cal L}$.
In general they are complex with the steady state eigenvector having an eigenvalue of zero.
The solution for this equation is:
$$
\vec Y(t) = e^{ M t} \vec Y(0).
$$
When $ M$ is diagonalizable, we can write $ M = T \Lambda T^{-1}$,
for a non-singular matrix $T$ and a diagonal matrix $\Lambda$, 
which has the eigenvalues $\lbrace \lambda_i \rbrace$ on the diagonal.  
Then we have
$ e^{ M t} = T e^{\Lambda t} T ^{-1} $,
with the diagonal matrix $e^{\Lambda t}$, which has the exponential of the eigenvalues, $e^{\lambda_i t}$, on its diagonal.
The resulting dynamics of expectation values of operators, as well as other correlation functions, 
follows a sum of decaying oscillatory exponentials.
The analytical form of such dynamics is:
\begin{equation}
\left\langle X(t) \right\rangle =
\sum_{k}\;d_k \exp[-i \omega_kt]\;,
\label{eq:har}
\end{equation}
where $-i \omega_k$, denoted as complex frequencies,
are the eigenvalues of $ M$, 
$d_k$ are the associated amplitudes, 
and both $\omega_k$ and $d_k$ can be complex.
The real part of the complex frequency $\omega_k$ represents the oscillation rate, 
while the imaginary part, $Im(\omega_k)\le0$ represents the decaying rate,

For special values of the system parameters
the spectrum of the non-hermitian matrix $ M$ is incomplete.
This is due to the coalescence of several eigenvectors, referred to as a non-hermitian degeneracy.
The difference between hermitian degeneracy and non-hermitian degeneracy is essential: 
In the hermitian degeneracy, several different orthogonal eigenvectors  are associated with the same eigenvalue.
In the case of non-hermitian degeneracy several eigenvectors coalesce to a single  eigenvector
\cite[Chapter 9]{moiseyev2011non}.
As a result, the matrix $ M$ is not diagonalizable.

The exponential of a non-diagonalizable matrix $ M$ can be expressed using its Jordan normal form: 
$ M = T J T ^{-1}$. 
Here, $J$ is a Jordan-blocks matrix which has (at least) one non-diagonal Jordan block;
$ J_i = \lambda_i I + N $, 
where $I$ is the identity  and $N$ has ones on its first upper off-diagonal. 
The exponential of $ M$ is expressed as $ e^{ M t} = T e^{J t} T ^{-1} $, 
with the block-diagonal matrix $e^{J t}$, 
which is composed from the exponential of the Jordan blocks $e^{J_i t}$. 
For non-hermitian degeneracy of an eigenvalue $\lambda_i$, 
the exponential of the block $J_i$  will have the form:
$
e^{J_i t} = e^{\lambda_i I t + N t} = e^{\lambda_it} e^{N t}.
$
The matrix $N$ is nilpotent and therefore the Taylor series of $e^{Nt}$ is finite, 
resulting in a polynomial in the matrix $Nt$.
This gives rise to a polynomial behaviour of the solution, 
and the dynamics of expectation values will have the analytical form of 
\begin{equation}
\left\langle X(t) \right\rangle =
\sum_{k}\sum_{\alpha=0}^{r_{k}}\; {d}_{k,\alpha} t^{\alpha} \exp[-i\omega_{k}^{(r_k)}t]\;,
\label{eq:exh}
\end{equation}
replacing the form of Eq. (\ref{eq:har}).
Here, $\omega_k^{(r_k)}$ denotes an eigenvalue with multiplicity of $r_k+1$.
Note that for non-degenerate eigenvalues, i.e. $r_k=0$, 
we have $d_{k,0}=d_k$ and $\omega_k^{(0)}= \omega_k$.
The difference in the analytic behaviour of the dynamics results in non-Lorentzian line shapes,
with higher order poles in the complex spectral domain.

The point in the spectrum where the  eigenvectors coalesce is known as an exceptional point ($EP$).
When two eigenvalues of the master equation coalesce into one, a second-order non-hermitian degeneracy is obtained.
We refer to it as  {\em EP2}, while a third-order non-hermitian degeneracy is denoted by {\em EP3}.

This study addresses the scenario of the dynamics of a system coupled to a bath.
The formalism is a reduced description of a tensor product of the system and the bath \cite{breuer2002theory,alicki2007qds}. 
The coupling to the bath introduces  dissipation and dephasing into the dynamics. 
The state is represented as a density operator in Liouville space,
and the dynamics is governed by the  L-GKS equation. 
The non hermitian properties of the dynamical generator ${\cal L}$ is caused by tracing out the bath degrees of freedom.
We employ the Heisenberg picture with a complete operator basis set in Liouville space.

Previous studies of the physics of {\em EP}s investigated the scenario of scattering resonances phenomena.
In that different scenario,
the non hermitian properties of the effective Hamiltonian are caused by the interaction between the discrete states
via the common continuum of the scattering states \cite{fano1961effects,rotter2009nonHermitian}.  
In those studies only coherent dynamics is considered and the dissipation and dephasing phenomena are absent. 

Examples for \emph{EP}s have been described in optics \cite{berry2004physics,uzdin2012scattering}, 
in atomic physics \cite{latinne1995laser,cartarius2007exceptional,uzdin2013effects,moiseyev2013sudden,rotter1999resonance,rotter2001continuum}, 
in electron-molecule collisions \cite{narevicius2000trapping},
superconductors \cite{rubinstein2007bifurcation}, 
quantum phase transitions in a system of interacting bosons \cite{cejnar2007coulomb}, 
electric field oscillations in microwave cavities \cite{dembowski2001experimental}, 
in PT-symmetric waveguides \cite{klaiman2008visualization}, 
and in mesoscopic physics \cite{muller2008EP,muller2009phaseLapses}. 

Recently, Wiersig suggested a method to enhance the sensitivity of detectors using exceptional points
\cite{Wiersig2014enhancingSensitivity}. 
Below we suggest to employ the exceptional points for the purpose of parameter estimation.

\section{Identifying the exceptional points and parameter estimation}

The analytical form of decaying exponentials, Eq. (\ref{eq:har}), 
is used in harmonic inversion methods to find the frequencies and amplitudes of the time series signal 
\cite{neuhauser1995FDM, mandelshtam2001FDM,main2000threeHI}. 
These frequencies and amplitudes can be employed to estimate the system parameters. 
If the sensitivity of the estimated frequencies is increased with respect to the system controls, 
the accuracy of the parameter estimation is enhanced.
Such sensitivity increase can be achieved using the special character of the dynamics at exceptional points. 

At exceptional points the analytical form includes also polynomials (Eq. (\ref{eq:exh})). 
Fuchs et al. showed that applying the standard harmonic inversion methods to a signal generated by Eq. (\ref{eq:exh})
leads to divergence of the amplitudes $d_k$.
An extended harmonic inversion method can fix the problem.
The divergence of the amplitudes $d_k$ at the vicinity of exceptional point can be used to locate them in the parameter space very accurately \cite{fuchs2014harmonic}. 
This is a consequence of the special non analytic character close to the {\em EP}
(Cf. in Chapter 9 in Ref. \cite{moiseyev2011non}).

Relying on the ability to accurately locate the \emph{EP}s in the parameter space, 
we suggest to use the \emph{EP}s for parameter estimation.
%In the following we suggest to use the exceptional points for parameter estimation, 
%relying on the ability to accurately locate them in the parameter space as discussed above. 
The procedure we suggest follows:
\begin{enumerate}
\item 
Accurately locate in the parameter space the desired exceptional point by iterating the following steps:
\begin{enumerate}
\item Perform the experiment to get a time series of an observable for example the polarization as a function of time. 
\item Obtain the characteristic frequencies and amplitudes of the signal using harmonic inversion methods.
\item In the parameter space, estimate the direction and distance to the \emph{EP} and determine new parameters for the next iteration.
\end{enumerate}
\item 
Invert the relations between the characteristic frequencies and the system parameters at the \emph{EP} to obtain the system parameters. 
\end{enumerate}

The accurate location of the exceptional points, followed by inverting the relations, will lead to accurate parameter estimation.

\section{Determination of the physical parameters in two level systems}
\subsection{The Bloch equation}
The Bloch equation describes the dynamics of  the three components of the nuclear spin,
$S_x$, $S_y$, and $S_z$,
under the influence of an external magnetic field, 
or a two-level atom in external electromagnetic field. 
In the rotating frame, we can write the equations in a matrix-vector notation:
\begin{eqnarray}
\frac{d}{dt}\left( \begin{array}{c} \tilde S_x\\ \tilde S_y\\ S_z \end{array} \right) 
=
\left( \begin{array}{ccc}
-\frac{1}{T_2}& \Delta & 0\\
-\Delta & -\frac{1}{T_2}& \epsilon\\
0 & -\epsilon & -\frac{1}{T_1}
\end{array} \right)
\left( \begin{array}{c} \tilde S_x\\ \tilde S_y\\ S_z \end{array} \right) 
+
\left( \begin{array}{c} 0\\ 0\\ \frac{1}{T_1} S_z^{0} \end{array} \right)
\label{eq:bloch_rotating},
\end{eqnarray}
with $T_1$ and $T_2$ as the dissipation and dephasing relaxation parameters, 
and the detuning from resonance $\Delta$ and the amplitude $\epsilon$ as the field parameters.
See details in \ref{sec:appendix:bloch}.

The Bloch equations can be derived from the L-GKS equation of the two-level system,  
with the effective rotating-frame Hamiltonian
%For the two level system, we define the vector space by the three traceless polarization operators $\Op S_x, \Op S_y,\Op S_z$
%and the identity $\Op I$. 
%The effective rotating-frame Hamiltonian of the system under a driving field
%with detuning $\Delta$ and driving frequency $\epsilon$ is:
$$
\Op H = \Delta \Op S_z + \epsilon \Op{\tilde{S}}_x,
$$
along with relaxation and dephasing terms.
See \ref{sec:appendix:bloch_deriv} for details.
%We use the polarization operators to form the  vector of basis operators: 
%$\vec{S}=(\Op S_{x},\Op S_{y},\Op S_{z})^{T}$.
%The equation of motion in the rotating frame is (see appendix \ref{sec:appendix:bloch_deriv}):
%$$
%\dot{\vec{S}} = \left(M - \gamma I \right) \left( \vec{S}- \vec{S}_{eq} \right),
%$$
%with $\Gamma$ the general relaxation coefficient, 
%$\gamma$ as the pure dephasing rate,
%and $ \vec{S}_{eq} $ the equilibrium polarization.
Reducing the number of parameters, 
the Master equation can be incorporated in the matrix:
\begin{equation}
\label{eq:M_bloch_matrix}
\text{M}=\left(
\begin{array}{ccc}
-\frac{\Gamma}{2} & \Delta & 0\\
-\Delta & -\frac{ \Gamma}{2} & \epsilon\\
0 & -\epsilon & \text{-\ensuremath{\Gamma}}
\end{array}
\right),
\end{equation}
with $\Gamma=\frac{3}{2}\frac{1}{T_1}-\frac{1}{T_2}$ as the general relaxation coefficient 
(See \ref{sec:appendix:bloch_deriv}).
%In this case the L-GKS master equation  is equivalent to the Bloch equation 
%(See appendix \ref{sec:appendix:bloch}). 
%This equation is a common form for TLS found in the literature \cite{fogli2007probing,gammelmark2014hidden,gammelmark2014fisher}.

The dynamics is determined by the exponential $e^{Mt}$, % in Eq. (\ref{eq:S_general_dynamics}).
which typically describes oscillating decaying signal, Cf. Eq. (\ref{eq:har}). 
Nevertheless, for specific parameters leading to  {\em EP} the dynamics is modified to include  polynomials, Cf. Eq. (\ref{eq:exh}).

\subsection{Exceptional points in the Bloch equation}

\begin{figure}[htbp]
\begin{center}
\includegraphics[scale=0.6]{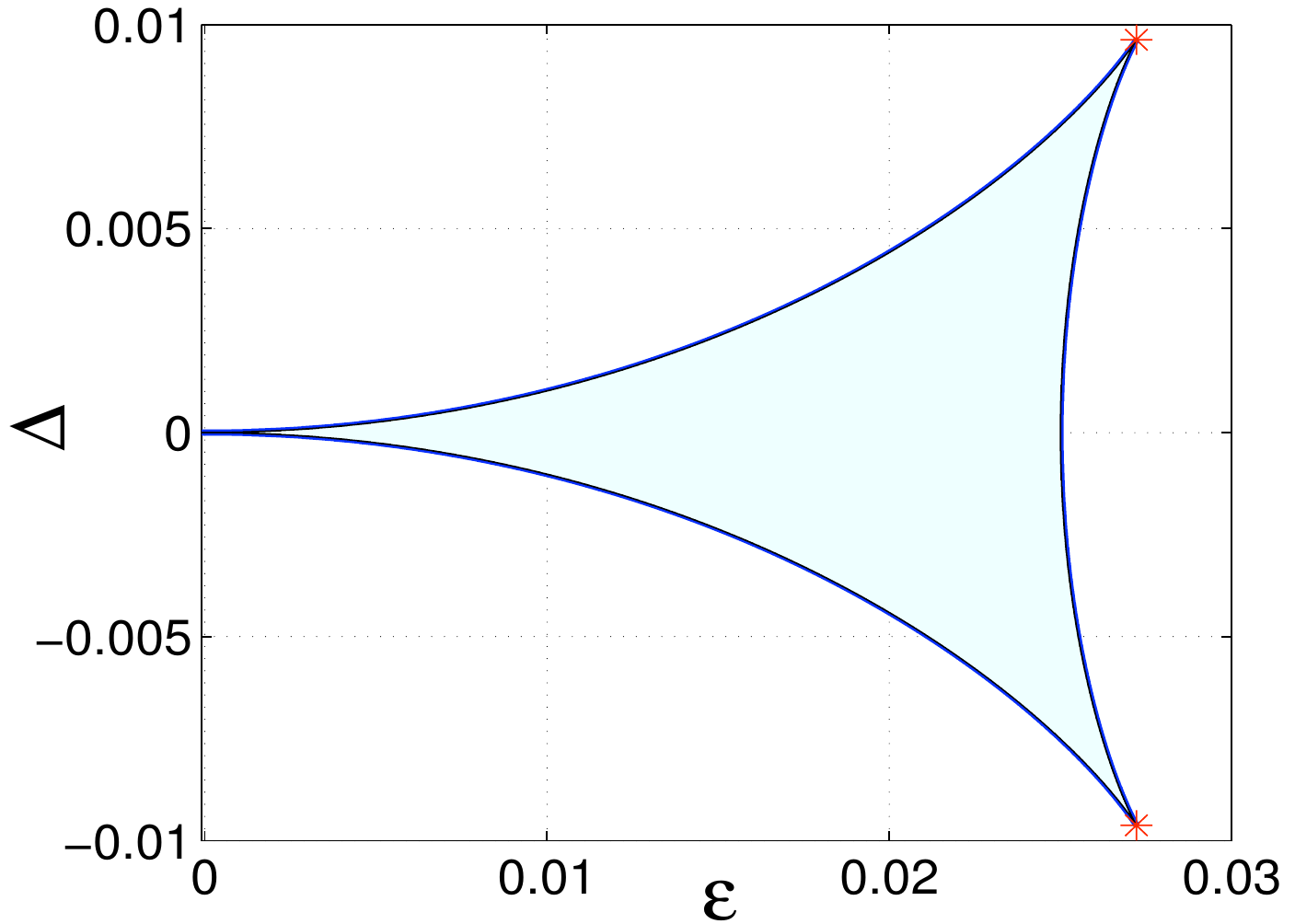}
\caption{
A map of the non-hermitian degeneracies of the eigenvalues of the matrix $ M$ of Eq. (\ref{eq:M_bloch_matrix}),
as a function of $\epsilon$ and $\Delta$,  for fixed $\Gamma=0.1$.
The lines represent second order  exceptional points ({\em EP2}).
The cusps,
where $\Delta=\pm \sqrt{1/108}\,\Gamma$,
$\epsilon=\sqrt{8/108}\, \Gamma$
(red asterisks),
are third order  exceptional point ({\em EP3}).
In the area inside the "triangle", marked with pale blue,
the eigenvalues of the matrix M are real.
The {\em EP2} curve distinguishes between points with real and complex eigenvalues.
}
\label{fig:full_ep_curve}
\end{center}
\end{figure}

The {\em EP}s are non-hermitian degeneracies in the matrix $M$ of Eq. (\ref{eq:M_bloch_matrix}). 
The task is to express the {\em EP}s using the parameters of this matrix.
Explicit derivations are presented in \ref{sec:appendix:eigenvalues}.
Non-hermitian degeneracies  of the eigenvalues \cite{moiseyev2011non}, {\em EP2}, occur when 
$$
\Gamma ^4 \Delta ^2+16 \left(\Delta ^2+\epsilon ^2\right)^3+\Gamma ^2 \left(8 \Delta ^4-20 \Delta ^2 \epsilon ^2-\epsilon ^4\right) = 0 .
$$
Figure \ref{fig:full_ep_curve} shows a map of {\em EP2} curve as a function of $\epsilon$ and $\Delta$ for fixed $\Gamma=0.1$.
Such figures were obtained in the study of analytical solutions for the Bloch equation
\cite{noh2010bloch,moroz2012unorthodox,noh2015bloch}.

A third order exceptional point, {\em EP3}, occurs  when 
$\Delta=\pm \sqrt{1/108}\,\Gamma$, $\epsilon=\sqrt{8/108}\,\Gamma$
(red asterisks in Figure \ref{fig:full_ep_curve}). 
These triple-degeneracies {\em EP3} occur twice,
and have a cusp-like behaviour, emerging from the {\em EP2}-curves, 
identifiable as a section through an elliptic umbilic catastrophe \cite{berry1979elliptic}.
This topology is also consistent with an analysis  
of non hermitian degeneracies in a two-parameters family of $3 \times 3$ matrices 
\cite{Mailybaev2006computation}. 
In very strong driving fields the matrix $ M$ will loose symmetry \cite{geva1995relaxation,szczygielski2013markovian}
maintaining the cusps but skewing the topology.

\subsection{\emph{EP} identification and parameter estimation}

We now describe the two steps of the method  for accurate determination the physical parameters. 
The first step is to identify the desired exceptional point using a sequence of measured time-dependent signals. 
The second step is to invert the relations and determine the system parameters.

\subsubsection{Identifying the second and third order exceptional points}
To identify the exceptional points we used time series of the polarization observable 
$S_z \equiv \langle  \Op S_z \rangle$, 
initially at the ground state.
We simulated the dynamics  with varying field parameters ($\epsilon$, $\Delta$) generating a time series
of polarization $S_z[n] = S_z (n \delta t)$. 
This signals served as the input for the harmonic inversion.

The parameters $\Delta$ and $\epsilon$ were tuned close to an {\em EP}.
%From equations (\ref{eq:har}) and (\ref{eq:exh}) we see that $S_z(t)$ can have one of the %following analytical forms,
%Depending on the system parameters:
%\begin{equation}
%\label{eq:SZ}
%\begin{array}{l}
Generically we should have
$$
S_z(t) = d_1 e^{-i\omega_1t} + d_2 e^{-i \omega_2 t}+ d_3 e^{-i \omega_3 t}, 
$$
%\text{ non-degenerate,}\\
but in the \emph{EP2} $(r_k=1)$ we get
$$
S_z(t) = d_1  e^{-i\omega_1 t}
+ \left( d_{2,0} + d_{2,1} t \right) e^{-i\omega_2^{(1)} t},
$$
%\text{  {\em EP2} } (r_k=1) \text{, }\\
and for \emph{EP3} $(r_k=2)$ 
$$
S_z(t) =  \left( d_{1,0} + d_{1,1} t +d_{1,2} t^2 \right) e^{-i\omega_1 ^{(2)} t}.
$$
(Cf. Eqs. (\ref{eq:har}) and (\ref{eq:exh})).
%\text{   {\em EP3} } (r_k=2).
%\begin{array}{rcll}
%S_z(t) & = & d_1 e^{-i\omega_1t} + d_2 e^{-i \omega_2 t}+ d_3 e^{-i \omega_3 t} ,
%& \text{non-degenerate}\\
%S_z(t) & = & d_1  e^{-i\omega_1^ t}
%+ \left( d_{2,0} + d_{2,1} t \right) e^{-i\omega_2^{(1)} t} ,
%& \text{{\em EP2} } (r_k=1)\\
%S_z(t) & = & \left( d_{1,0} + d_{1,1} t +d_{1,2} t^2 \right) e^{-i\omega_1 ^{(2)} t} ,
%& \text{{\em EP3} } (r_k=2).
%\end{array}
%\end{equation}
We located suspected \emph{EP}s by identifying possible degeneracies of the assigned frequencies $\omega_k$.
As stated earlier, applying standard harmonic inversion methods for the time series generated by a non-diagonalizable matrix,
leads to divergence of the amplitudes $d_k$ \cite{fuchs2014harmonic}.
This divergence can be used to locate the exceptional points accurately.
A verification can be obtained by using the extended harmonic inversion method. 

\begin{figure}[htbp]
\begin{center}
\includegraphics[scale=0.4]{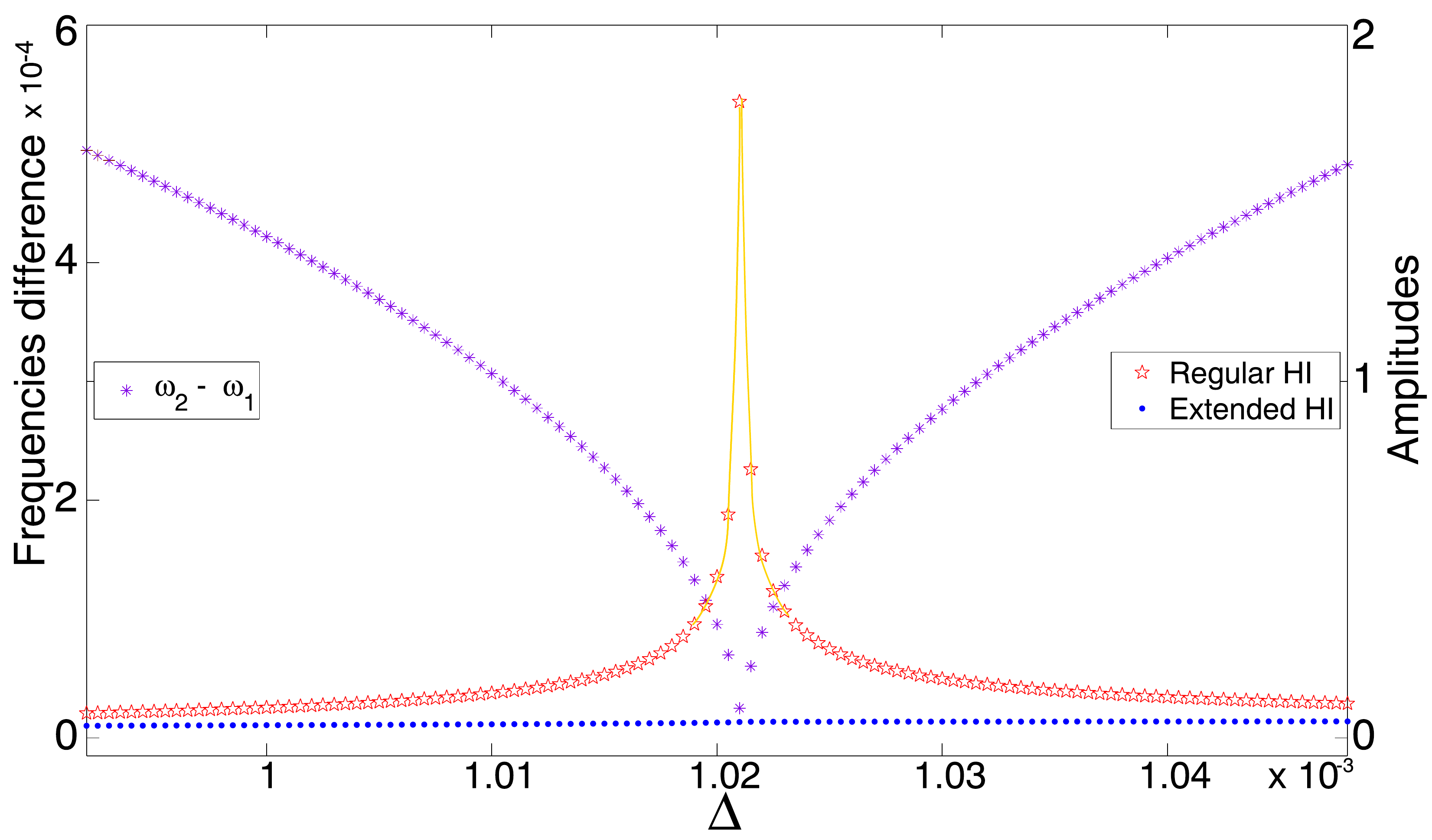}
\caption{
Identifying an {\em EP2} for $\Gamma=0.1$ and $\epsilon=0.01$. 
The left y-axis (purple asterisks) shows the absolute value of the difference between the frequencies, $|\omega_2-\omega_1|$,
versus the detuning $\Delta$.
The non-hermitian degeneracy point is located with high resolution.
The right y-axis shows the corresponding amplitude, 
obtained by the regular harmonic inversion method $|d_1|$ 
(red stars),
and by the extended method  $|d_{1,0}|$ (blue points).
The diverging behaviour of $|d_1|$ indicates that the degeneracy is an \emph{EP}.
}
\label{fig:harminv_eps0p01}
\end{center}
\end{figure}

This procedure was employed to identify an \emph{EP2} for fixed $\Gamma=0.1$ and $\epsilon=0.01$, 
with varying $\Delta$.
The purple asterisks at Fig. \ref{fig:harminv_eps0p01} displays the absolute value of the difference between the frequencies 
$|\omega_2-\omega_1|$, obtained by the harmonic inversion for each parameter set.
The degeneracy point is clearly observed.
The diverging behaviour of the amplitudes is shown in red stars.
It is consistent with the degeneracy of the frequencies.
The {\em EP2} is located at $\Delta=1.021\times10^{-3}$, consistent with the  prediction.
Using a finer mesh of sampling points the \emph{EP}
can be identified with a resolution exceeding  $0.5\times10^{-9}$.

\begin{figure}[htbp]
\begin{center}
\includegraphics[scale=0.65]{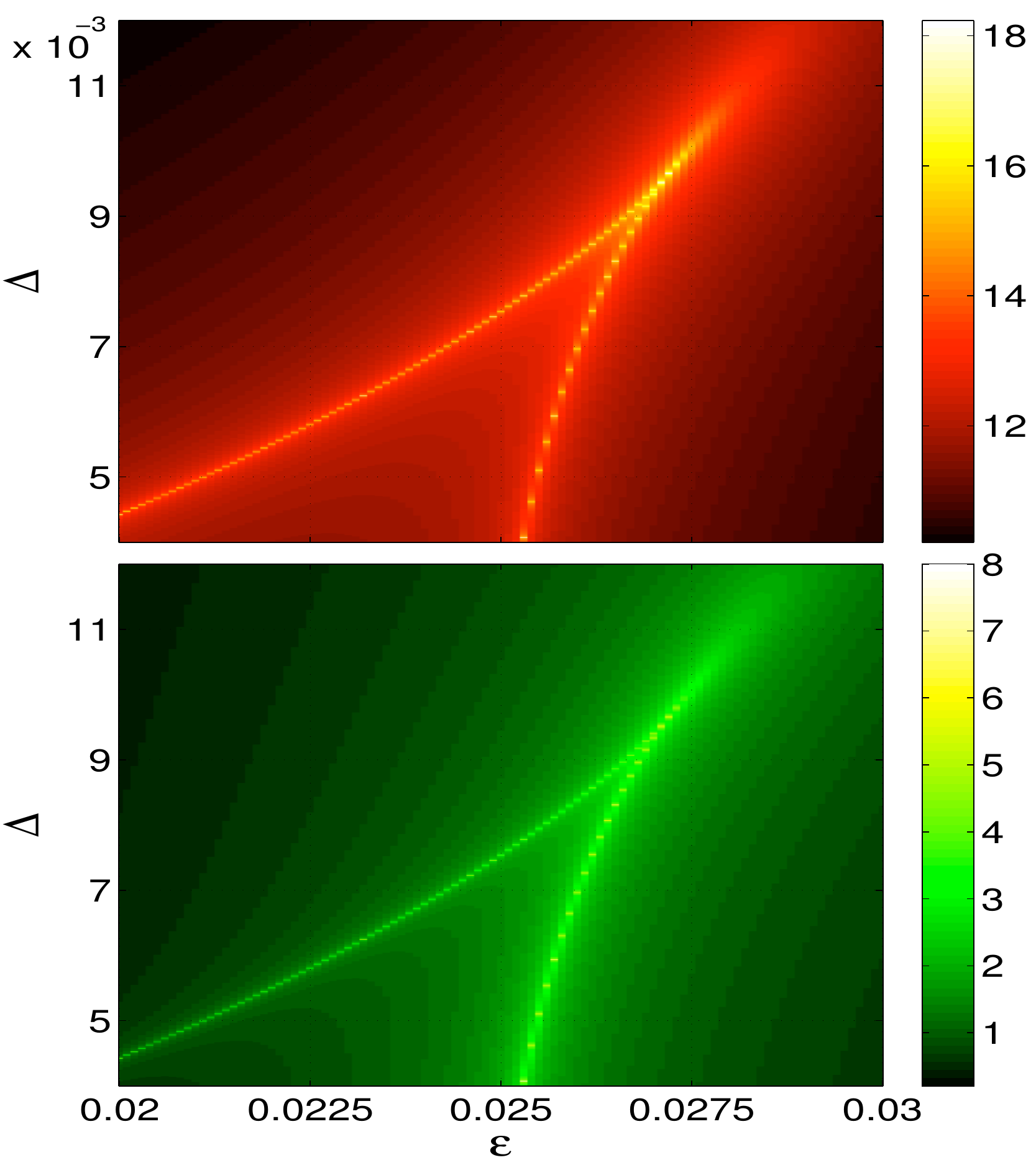}
\caption{
Identifying the triple exceptional point {\em EP3}.
The upper panel shows the 2-D function
$F(\Delta,\epsilon,\Gamma)
%= \log \left(  \left| \frac{1}{( \omega_1- \omega_2)} \frac{1}{( \omega_2- \omega_3)} \frac{1}{( \omega_3- \omega_1)} \right| \right)
$
presented in the text. 
%where $ \omega_1$, $ \omega_2$ and $ \omega_3$ are the complex frequencies
%obtained from harmonic inversion Eq. (\ref{eq:har}).
The highest point corresponds to the triple-{\em EP} point {\em EP3}.
The lower panel shows the sum of the absolute values of the amplitudes,
which were calculated by the regular harmonic inversion method.
}
\label{fig:harminv_triple_point_degen_amps}
\end{center}
\end{figure}

%{\bf Identifying the {\em EP3} by varying $\epsilon$ and $\Delta$}.
The \emph{EP3} was identified by a 2-D search performed by 
varying  $\epsilon$ and $\Delta$, for fixed $\Gamma=0.1$.
We searched for the degeneracies of the three eigenvalues
by employing the 2-D function
\begin{equation}
\label{eq:F_w_function}
F(\Delta,\epsilon,\Gamma)= \log \left(  \left|
\frac{1}{( \omega_1- \omega_2)}
\frac{1}{( \omega_2- \omega_3)}
\frac{1}{( \omega_3- \omega_1)}
\right| \right),
\end{equation}
which should diverges at the {\em EP} curve. 
Numerically, we get high values at this curve, with highest values obtained at the {\em EP3}.
The upper panel of Figure \ref{fig:harminv_triple_point_degen_amps} shows the sharp curve of peaks  following the curve of exceptional points.
The highest point on the merging two ridges is the {\em EP3}.
The lower panel of Figure \ref{fig:harminv_triple_point_degen_amps} shows the sum of the absolute values of the amplitudes,
calculated by the standard harmonic inversion. 
The curve of the exceptional points is clearly identified.

%A very fine resolution in the $\Delta$-$\epsilon$ space was crucial  for exceptional points detection.
%The entire structure of two peaks in the amplitude is found in a section of length $10^{-5}$ in $\Delta$,
%while the "triangle" of exceptional points is located on an area of $2\times10^{-2}$ in $\Delta$ (see Figure \ref{fig:full_ep_curve} above).

Refining the search leads to very high resolution,
and the {\em EP3} can be identified with a high accuracy, approaching 
the theoretical values of
$\Delta=\sqrt{1/108}\,  \Gamma$,
$\epsilon=\sqrt{8/108}\,  \Gamma$.

An efficient algorithm to identify the \emph{EP3} is demonstrated based on
a two-dimensional search in the parameter space of $\Delta$ and $\epsilon$.
This procedure enables the experimentalists to identify accurately the laser parameters for which the \emph{EP3} is obtained.
We use the maximum of the function  Eq. (\ref{eq:F_w_function}) as the objective leading to \emph{EP3}.
%This function diverges at the degeneracies. 
%Numerically, it gets high values at this curve, with highest values obtained at the {\em EP3}.

Evaluating the function at each desired point in the parameter space include the following steps:
\begin{enumerate}
	\item \textbf{Time series}: Obtain a time series of the polarization by performing the experiment or the numerical simulation.
	\item \textbf{Frequencies}: Calculate the frequencies from the time series by harmonic inversion.
	\item \textbf{Function evaluation}: Evaluate the function $F(\Delta,\epsilon,\Gamma)$ from the calculated frequencies.
\end{enumerate}

\begin{figure}[htbp]
	\begin{center}
		\includegraphics[scale=0.45]{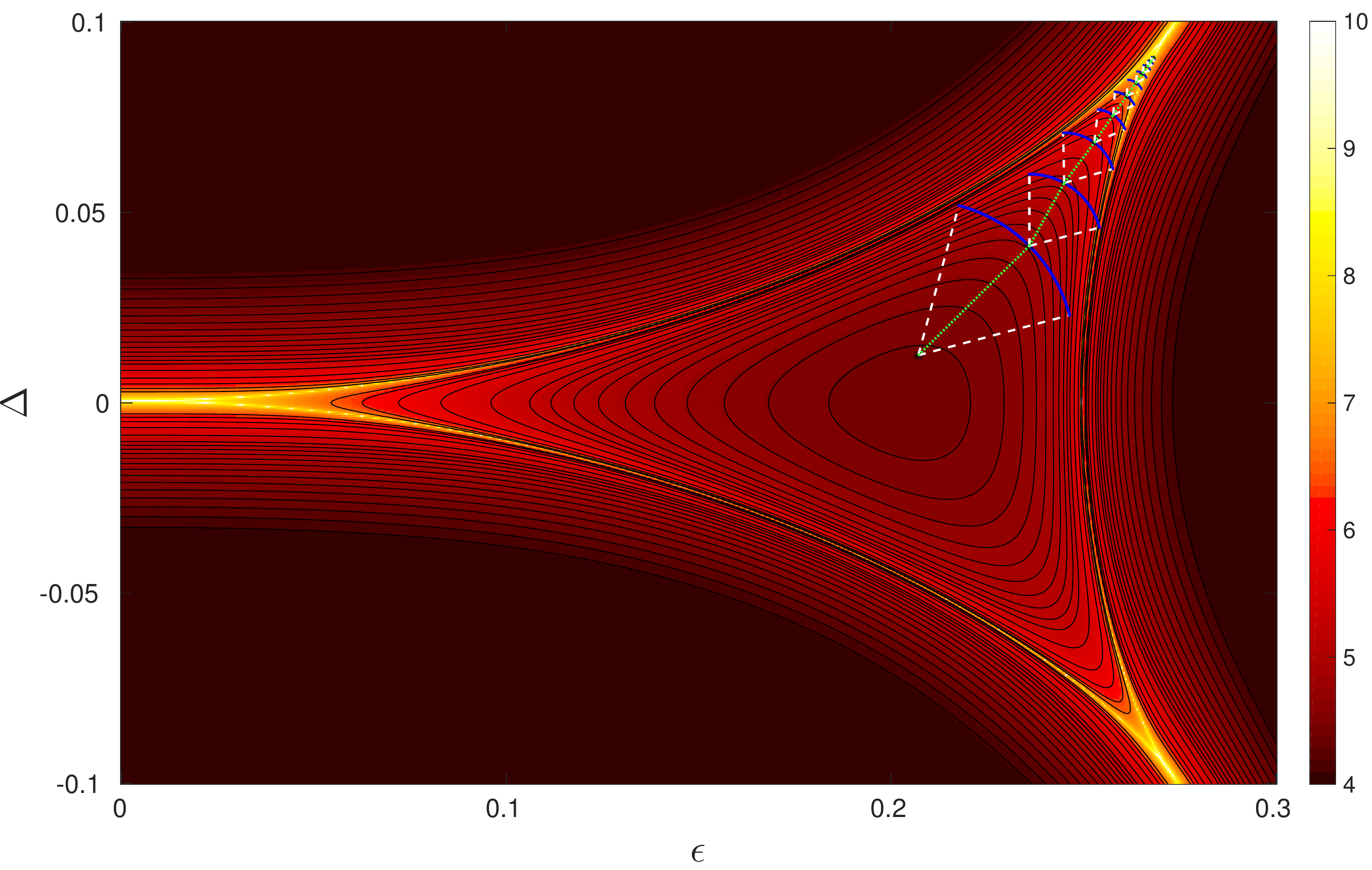}
		\caption{
			A sketch of the iterations progress in the "valley ascend" method.
			The collors on the background and the black contour lines represent the function 
			$F(\Delta,\epsilon,\Gamma)$ of Eq. (\ref{eq:F_w_function}). 
			In each iteration we plotted with blue line the circular arc on which we searched for the minimum.
			The black asterisks show these minima, 
			which form the curve, plotted with a dashed light green line, 
			that "climbs" in the valley of the objective function.
		}
		\label{fig:climb_valley}
	\end{center}
\end{figure}

Standard search methods can stagnate due to the high values at the \emph{EP2} curve. 
Another difficulty is the cusp behaviour of the \emph{EP2} curve close to the \emph{EP3}.
To overcome these diffculties we implemented a "climbing the valley" procedure: 
Staying on the valley of the local minima ensures the search overcomes the stagnation due to the  \emph{EP2} curve. 
The procedure follows:

\begin{enumerate}
	\item \textbf{Preliminary step - initial point}: 
	\begin{enumerate}
		\item Locate points inside the triangle-like \emph{EP} curve (Cf.  Fig. \ref{fig:climb_valley}). 
		The inner area of the curve is characterized by real-only eigenvalues. 
		\item Perform a 1-D search to find a minimum on a straight line.
	\end{enumerate}
	
	\item \textbf{Valley ascend}: Each iteration ascends up the valley to a  point  with higher value of the function $F$. 
	This is done by finding a minimum on the circular arc that is centred at the current point, enclosed by two radii. 
	The angles of these radii can be predefined or defined on each iteration.
	We perform the following steps:
	\begin{enumerate}
		\item \emph{Determining the angular range}. 
		Predefined or from the previous iterations.
		\item \emph{Determining the radius}. 
		The radius is the distance from the current point to nearest point on the \emph{EP2} curve that is in the angular range.
		\item \emph{Finding the next point}.
		Performing a 1-D search on the circular arc that is defined by the angular range and the radius
		(Cf. blue arc in Fig. \ref{fig:climb_valley}). 
		The point for the next iteration is the point on the arc with the minimal value of $F$
		(Cf. end of green line in Fig. \ref{fig:climb_valley}).
	\end{enumerate}
	
	These steps converge to the desired \emph{EP3} point.
	Figure \ref{fig:climb_valley} demonstrates the progress in the "valley ascend" method with a few iterations.
	
\end{enumerate}

The \emph{Valley ascend} method presented above is a generic method, 
and can be used also for searching higher order degeneracies in other systems.
For the Bloch equation case, where the generating matrix, Eq. (\ref{eq:M_bloch_matrix}), is a $3\times3$ matrix, 
the \emph{EP3} is the point where the characteristic polynomial
%The  is the triple degeneracy of the matrix eigenvalues,
%and therefore it is the point where the characteristic polynomial
\begin{equation}
\label{eq:p_w_polynom}
P_{_{\Delta,\epsilon,\Gamma}}(\omega) = 
( \omega- \omega_1)
( \omega- \omega_2)
( \omega- \omega_3)
\end{equation}
has roots with multiplicity of 3. 
Therefore we can use the special properties of the cubic equation and perform a regular root search.
We define $r$, $s$ and $t$ as the coefficient of the polynomial $P_{_{\Delta,\epsilon,\Gamma}}(\omega)$ defined in Eq. (\ref{eq:p_w_polynom}):
\begin{equation}
\label{eq:r_s_t}
( \omega- \omega_1)
( \omega- \omega_2)
( \omega- \omega_3)
 = \omega^3 + r \omega^2 + s \omega + t.
\end{equation}
We define the functions
\begin{equation}
\begin{array}{rcl}
p(\Delta,\epsilon,\Gamma) & = & s-\frac{1}{3}r^{2} \\
q(\Delta,\epsilon,\Gamma) & = & \frac{2}{27}r^{3}-\frac{1}{3}rs+t,
\end{array}
\end{equation}
and perform a 2D conventional root search. 
The point in the parameter space where these two functions vanish is point where the three eigenvalues are degenerate.
We have applied this method using standard method of 2D root search obtaining high accurate values of the $EP3$.

\subsubsection{Physical parameter estimation from the value at the exceptional point}

For the two-level-system the system parameters are the frequency $\omega_s$ associated with the energy gap, 
the general decay rate $\Gamma$ and the dipole strength $\mu$.
The external experimentally controlled parameters are the driving frequency $\nu$ and the power amplitude $\cal E$.
The parameters of Eq. (\ref{eq:M_bloch_matrix}) can be related with $\epsilon = \mu {\cal E}$  and $\Delta =\omega_s-\nu$.
One would like to estimate the system parameters from experiments.
After locating accurately the \emph{EP}, 
we can determine the parameters by inverting the relations between the eigenvalues and the system parameters.

To obtain high accuracy, we used the identification the triple-degeneracy point {\em EP3} presented above, 
so both parameters - $\Delta$ and $\epsilon$ - are located accurately. 
The accurate location of $\Delta$ and $\epsilon$ makes the parameter estimation very robust 
to uncertainties in the location of the exceptional points.
This is a consequence of the special non analytic character close to the \emph{EP3} 
(Cf. \ref{sec:appendix:non_analytic}).
Therefore, the system parameters 
$\Gamma$, $\omega_s$ and $\mu$ can be determined to a high degree of accuracy at this point.
From the eigenvalues of the matrix $\text{M}$ in Eq. (\ref{eq:M_bloch_matrix}) we get
$ \Gamma = \frac{i}{2} \left(\omega_1+\omega_2+\omega_3 \right) $.
To obtain $\epsilon$ and $\Delta$ one has to invert non-linear relations 
(see \ref{sec:appendix:eigenvalues}).
At the \emph{EP3}, the inversion becomes: $\textstyle \omega_s= \nu +\sqrt{1/108}\, \Gamma$, 
$\mu = \sqrt{8/108}\, \Gamma/ {\cal E}$.

\subsubsection{Noise sensitivity}
Parameters estimation naturally raises the issue of sensitivity to noisy experimental data.
The noise sensitivity will be determined by the method of harmonic inversion. 
If the sampling periods have high accuracy then the time series can be shown to have an underlying Hamiltonian
generator. 
This is the basis for linear methods, such as the filter diagonalization (FD) 
\cite{neuhauser1995FDM,mandelshtam2001FDM}. 
The noise in these methods results in normally distributed underlying matrices, 
and the model displays monotonous behaviour with respect to the noise.
This was verified analytically and by means of simulations In Ref. \cite{Benko2008noisyFDM}.
As a result sufficient averaging will eliminate the noise.
Practical implementations require further analysis with evidence of nonlinear effects of noise. 
For example, Mandelshtam et. al. analysed the noise-sensitivity of the FD in the context of NMR experiments 
\cite{hu199876FDM_noise,Celik2010FDM_sensitivity} 
and Fourier transform mass spectrometry \cite{martini2014mass_spectrometry}.
For some other methods, a noise reduction technique was proposed in Ref. \cite{main2000threeHI}.

%Evaluating the intrinsic parameters of the system from experiment is a major task. 
%The non-unitary characteristics of open systems can assist in this task.
%M{\o}lmer et al. \cite{gammelmark2014fisher,kiilerich2014estimation} 
%suggested a method of parameter estimation for a two-level system based on the time correlation function of single photon emission.

\section{Discussion}
Bloch's equation has become the template for the dynamics of open quantum systems. 
Such systems typically decohere with a dynamical signature of decaying oscillatory motion. 
It is therefore surprising that the existence of non hermitian degeneracies has been overlooked. 
Our finding of an intricate manifold of double degeneracies {\em EP2} and
triple degeneracies {\em EP3} in the elementary TLS template suggests that 
any quantum dynamics described by the L-GKS generator \cite{lindblad1976generators,gorini1976completely}
will exhibit a manifold of exceptional points.

Non hermitian degeneracies of the {\em EP} have a subtle influence on the dynamics. 
The hallmark of {\em EP} dynamics is a polynomial component in the decay leading to non-Lorentzian lineshapes.
We suggest an experimental procedure to identify the {\em EP} in Bloch systems, 
using harmonic inversion of the polarization time series.
The sensitivity of harmonic inversion in the neighbourhood of an  {\em EP} 
enables us to accurately locate the {\em EP}, 
and therefore allows us to determine the system parameters: 
the energy gap $\omega_s$, 
the dipole transition moment $\mu$, 
and the decoherence rate $\Gamma$. 

This study is only the first step in establishing parameter estimation via exceptional points.
A generalization to larger Liouville spaces is  under study for atomic spectroscopy.
Under the influence of driving fields and due to spontaneous emission, 
atoms and ions can have a structure of N-level system with relaxation.
In these systems we expect non-hermitian degeneracy of high order.
The structure of the exceptional points in these systems can be used for estimating the energy differences, the lifetimes, and branching ratios.
work in this direction is in progress \cite{morag2015EPatomic}.

Many quantum systems are open and their dynamics has dissipative nature, 
which is described well by the L-GKS equation.
Therefore we expect to find exceptional points in many quantum systems. 
Under the appropriate circumstances these \emph{EP}s can be used for accurate parameter estimation.

\subsection*{Aknowledgements}
We thank Ido Schaefer, Amikam Levy, and Raam Uzdin for fruitful discussions.
We thank Jacob Fuchs and J\"org Main for assisting with the extended harmonic inversion method.
We thank the referee for proposing the root search for the \emph{EP3}.
Work supported by the Israel Science Foundation   Grants No. 2244/14 and No. 298/11  and  by I-Core: the Israeli Excellence Center ``Circle of Light''.

\appendix

\section{Bloch equations}
\label{sec:appendix:bloch}
The Bloch equation describes the dynamics of  the three components of the nuclear spin,
$S_x$, $S_y$, and $S_z$,
under the influence of an external magnetic field $\vec{H}$.
The equations as appear in Bloch's original paper (\cite{bloch1946nuclear}, Eq. 38) are

\begin{eqnarray}
\begin{array}{rcl}
\dot{S_x} & = & \gamma \left( S_y H_z - S_z H_y \right) - \frac{1}{T_2} S_x \\
\dot{S_y} & = & \gamma \left( S_z H_x - S_x H_z \right) - \frac{1}{T_2} S_y \\
\dot{S_z} & = & \gamma \left( S_x H_y - S_y H_x \right) - \frac{1}{T_1} \left( S_z-S_z^{0} \right).
\end{array}
\label{eq:original_bloch_xyz}
\end{eqnarray}
$T_1$ and $T_2 $ are two relaxation parameters 
(the pure dephasing rate $\frac{1}{T_2^*}$ is related by $\frac{1}{T_2 }= \frac{1}{2 T_1}+\frac{1}{T_2^*}$), 
$\gamma$ is the gyromagnetic ratio, 
and $S_z^{0}$ is the equilibrium value of $S_z$ under the influence of constant external magnetic field $H_z = H_0$.
These equations can be recast in a matrix-vector notation:
\begin{eqnarray}
\frac{d}{dt}\left( \begin{array}{c} S_x\\ S_y\\ S_z \end{array} \right) 
=
\left( \begin{array}{ccc}
-\frac{1}{T_2}& \gamma H_z & -\gamma H_y \\
-\gamma H_z & -\frac{1}{T_2}& \gamma H_x \\
\gamma H_y & -\gamma H_x & -\frac{1}{T_1}
\end{array} \right)
\left( \begin{array}{c} S_x\\ S_y\\ S_z \end{array} \right) 
+
\left( \begin{array}{c} 0\\ 0\\ \frac{1}{T_1} S_z^{0} \end{array} \right)
\label{eq:original_bloch_matrix}.
\end{eqnarray}

For an external field $\vec{H}$ with the components 
$H_x = H_1 \cos \omega t$ , 
$H_y = - H_1 \sin \omega t$,
$H_z = H_0$, 
we define the rotating frame:
\begin{eqnarray}
\begin{array}{rcl}
S_x & = & \phantom{-} \tilde{S_x} \cos \omega t - \tilde{S_y} \sin \omega t \\
S_y & = & -\tilde{S_x} \sin \omega t - \tilde{S_y} \cos \omega t \\.
\end{array}
\end{eqnarray}

With the notations $\epsilon=\gamma H_1$ and $\Delta=\gamma H_0-\omega$ 
we have (see also \cite{Brewer1983bloch}):
\begin{eqnarray}
\frac{d}{dt}\left( \begin{array}{c} \tilde S_x\\ \tilde S_y\\ S_z \end{array} \right) 
=
\left( \begin{array}{ccc}
-\frac{1}{T_2}& \Delta & 0\\
-\Delta & -\frac{1}{T_2}& \epsilon\\
0 & -\epsilon & -\frac{1}{T_1}
\end{array} \right)
\left( \begin{array}{c} \tilde S_x\\ \tilde S_y\\ S_z \end{array} \right) 
+
\left( \begin{array}{c} 0\\ 0\\ \frac{1}{T_1} S_z^{0} \end{array} \right)
\label{eq:appendix:bloch_rotating}.
\end{eqnarray}

These equations also describe, in the dipole approximation, a two-level atom in external electromagnetic field.
In this case, the system parameters are the the unperturbed frequency of the system $\omega_s$, and the dipole strength $\mu$.
The external experimentally controlled parameters are the driving frequency $\nu$ and the power amplitude $\cal E$.
The parameters of Eq. (\ref{eq:appendix:bloch_rotating}) are related with $\epsilon = \mu {\cal E}$  and $\Delta =\omega_s-\nu$.
In the absence of dissipation the eigenvalues of the matrix are pure imaginary, 
and the dynamics is a free precession of the polarization vector characterized by the  Rabi frequency: $\Omega=\sqrt{\epsilon^2+\Delta^2}$.
When dissipation is present  the eigenvalues of the homogeneous part
of Eq. (\ref{eq:appendix:bloch_rotating}) become complex, reflecting a decaying oscillation dynamics leading asymptotically to a steady state.

\section{Derivation of the Bloch equation from the L-GKS equation}
\label{sec:appendix:bloch_deriv}
In the Heisenberg representation the L-GKS generator becomes:
\begin{equation}
\frac{d}{dt}\Op X~= 
~\frac{\partial \Op X }{\partial t}
+ i\left[\hat{\boldsymbol{\text{H}}},\Op X \right]
+ \sum_{k}\left(
\hat{\boldsymbol{\text{V}}}_{k}^{\dagger} \Op X\hat{\boldsymbol{\text{V}}}_{k}
- \frac{1}{2}\left\{ \hat{\boldsymbol{\text{V}}}_{k}^{\dagger}\hat{\boldsymbol{\text{V}}}_{k},\Op X \right\} 
\right).
\label{eq:appendix:LGKS}
\end{equation}
where $\Op X$ is an arbitrary operator. 
The Hamiltonian $\Op H$ is hermitian  and $\Op V$ is defined to operate in the Hilbert space of the system.
The curly brackets denote an anti commutator.
The set of operators $\lbrace \Op X \rbrace$ supports a Hilbert space construction, 
with the scalar product defined as:
$
\left( \Op{X}_1 , \Op{X}_2 \right) \equiv 
tr \left\lbrace \Op{X}_1^\dagger \Op{X}_2\right\rbrace
$.

For two-level system, the effective rotating-frame Hamiltonian under a driving field
with detuning $\Delta$ and driving frequency $\epsilon$ is:
\begin{equation}
\Op H = \Delta \Op S_z + \epsilon \boldsymbol{\mathrm{\tilde{S}}}_x 
\end{equation}
The two-level-system L-GKS equation for an operator $\Op X$ with relaxation and pure dephasing becomes
\begin{equation}
\begin{array}{rcl}
\frac{d}{dt} \Op X & = & i \; [ \Op H,\Op X] \\
&& + \,  \kappa_- \left(\Op S_+ \Op X \Op S_- -\frac{1}{2}\{\Op S_+\Op S_-, \Op X\}\right ) \\
&& + \,  \kappa_+ \left(\Op S_- \Op X \Op S_+ -\frac{1}{2}\{\Op S_-\Op S_+, \Op X\}\right ) \\ 
&& - \, \gamma \; [\Op S_z , [\Op S_z, \Op X]]
\label{eq:tlslgks}
\end{array}
\end{equation}
where $\kappa_{\pm}$ are kinetic coefficients, $\kappa_+/\kappa_- =\exp( -\hbar \omega /k_B T)$, and $\gamma$ is the pure dephasing rate \cite{agarwal1970master,emch1979standard}.

To rephrase the equation in a matrix-vector notation,
We use the polarization operators and the identity matrix to form the vector of basis operators: 
$\vec{S}^\prime =\left(\boldsymbol{\mathrm{\tilde{S}}}_x,
\boldsymbol{\mathrm{\tilde{S}}}_y,\Op S_{z}, \Op I \right)^{T}$.
Then Eq. (\ref{eq:tlslgks}) can be written as 
$ \dot{\vec{S}^\prime} = M^\prime \vec{S}^\prime $,
with an appropriate $4 \times 4$ matrix $M^\prime$.
We can reduce the dimensions by writing an inhomogeneous equation 
for the 3-component vector 
$\vec{S} = \left(\boldsymbol{\mathrm{\tilde{S}}}_x,
\boldsymbol{\mathrm{\tilde{S}}}_y,\Op S_{z} \right)^{T}$:

\begin{equation}
\dot{\vec{S}}=\left(M - \gamma I \right) \left( \vec{S}- \vec{S}_{eq} \right),
\label{eq:tls_matrix_notation}
\end{equation}

with $\Gamma=\kappa_- +\kappa_+-\gamma$, 
$I$ as the $3 \times 3$ identity matrix,
$ \vec{S}_{eq} $ that fulfills $(\gamma I - M ) \vec{S}_{eq} = \left(0,0,(\kappa_+ -\kappa_-)\Op I \right)^{T} $
%$\vec{e}_{3}=\left(0,0,\Op I \right)^{T}$, 
and the matrix:

\begin{equation}
\label{eq:appendix:M_bloch_matrix}
\text{M}=\left(
\begin{array}{ccc}
-\frac{\Gamma}{2} & \Delta & 0\\
-\Delta & -\frac{ \Gamma}{2} & \epsilon\\
0 & -\epsilon & \text{-\ensuremath{\Gamma}}
\end{array}
\right).
\end{equation}

Eq. (\ref{eq:tls_matrix_notation}) can be merged with the Bloch's equation (\ref{eq:appendix:bloch_rotating})
where $\frac{1}{T_1}=\kappa_+ +\kappa_-$ and $\frac{1}{T_2}=\gamma+\frac{1}{2}(\kappa_+ +\kappa_-)$.

The general solution for this equation is:
\begin{equation}
\label{eq:9}
\vec{S}(t)=e^{ -\gamma t}e^{\text{M}t}(\vec{S}_0-\vec{S}_{eq})+\vec{S}_{eq},
\end{equation} 
with $\vec{S}_{0}=\vec{S}(0)$.

The master equation Eq. (\ref{eq:tlslgks}) is a common form for TLS found in the literature \cite{fogli2007probing,gammelmark2014hidden,gammelmark2014fisher}.
Eq. (\ref{eq:appendix:M_bloch_matrix}) which determines the {\em EP} interpolates between
two extreme cases. The first is associated with  spontaneous emission, then $\Gamma=\kappa_-$.
The second is a hot singular bath dominated by pure dephasing, then $\Gamma=-\gamma$.

\section{Eigenvalues of the matrix M}
\label{sec:appendix:eigenvalues}
The task is to find the eigenvalues of the generator matrix (\ref{eq:M_bloch_matrix}).

We first define the variables:
\begin{equation}
\begin{array}{rcr}
Y & = & 12\Delta^{2}+12\epsilon^{2}-\text{\ensuremath{ \Gamma}}^{2}\\
X & = & -36\Delta^{2}+18\epsilon^{2}- \Gamma^{2}
\end{array}
\end{equation}
We also define:
\begin{equation}
\begin{array}{rcl}
W & = & \sqrt{ \Gamma^{2}X^{2}+Y^{3}}\\
& = & \left( \Gamma ^4 \Delta ^2+16 \left(\Delta ^2+\epsilon ^2\right)^3+\Gamma ^2 \left(8 \Delta ^4-20 \Delta ^2 \epsilon ^2-\epsilon ^4\right) \right)^{1/2}.
\end{array}
\label{eq:w}
\end{equation}

With these definitions  the eigenvalues of Eq. (\ref{eq:M_bloch_matrix}) become:
\begin{equation}
\label{eq:eigenvalues_M}
\begin{array}{ccc}
m_{1} & = & -\frac{2}{3} \Gamma+\frac{1}{6}\left(\left(W+ \Gamma X\right)^{1/3}-\frac{Y}{\left(W+\Gamma X\right)^{1/3}}\right)\\
m_{2} & = & -\frac{2}{3} \Gamma+\frac{1}{6}\left(e^{i\frac{2}{3}\pi}\left(W+\Gamma X\right)^{1/3} + e^{i\frac{1}{3}\pi}\frac{Y}{\left(W+\Gamma X\right)^{1/3}}\right)\\
m_{3} & = & -\frac{2}{3} \Gamma+\frac{1}{6}\left(e^{-i\frac{2}{3}\pi}\left(W+ \Gamma X\right)^{1/3} + e^{-i\frac{1}{3}\pi}\frac{Y}{\left(W+ \Gamma X\right)^{1/3}}\right).
\end{array}
\end{equation}
For real $W$ (i.e. for $\Gamma^{2}X^{2}+Y^{3} \ge 0$) all eigenvalues are real.
For $\Gamma^{2}X^{2}+Y^{3}<0$, $W$ is complex,  and  two of the eigenvalues are complex (complex conjugate to each other).

Non-hermitian degeneracies of the eigenvalues
occur when $W$ vanishes.
In such cases the second and third eigenvalues are degenerated,
leading to {\em EP2}.
A third order exceptional point, {\em EP3}, occurs for $X=Y=0$.
This happens when 
$\Delta=\pm \sqrt{1/108}\,\Gamma$, $\epsilon=\sqrt{8/108}\,\Gamma$. 
These triple-degeneracies {\em EP3} occur twice,
and have a cusp-like behaviour, emerging from the {\em EP2}-curves, identifiable as an elliptic umbilic catastrophe \cite{berry1979elliptic}.
This topology is also consistent with  the analysis 
of non hermitian degeneracies of a two-parameters family of $3 \times 3$ matrices,
done by Mailybaev \cite{Mailybaev2006computation}. 
In very strong driving fields the matrix $ M$ will loose symmetry \cite{geva1995relaxation,szczygielski2013markovian}
maintaining the cusps but skewing the topology.

\section{non analytic character close to the {\em EP3}}
\label{sec:appendix:non_analytic}
There is a special non analytic character close to the {\em EP3}:
When $\nu\to \nu^{EP3}$ and ${\cal E}\to{\cal E}^{EP3}$ 
then the three frequencies obtained by the standard harmonic inversion coalesce, 
leading to a branch point 
(Cf. Chapter 9 in Ref. \cite{moiseyev2011non}):
\begin{equation}
\omega_{k=1,2,3} =\omega_1^{(2)}+ e^{i\frac{2\pi}{3}}\left[\alpha_k(\nu-\nu^{EP3})+\beta_k({\cal E}-{\cal E}^{EP3})\right]^{\frac{1}{3}}
\end{equation}
where $\alpha_k $ and $\beta_k$ are parameters.
At the {\em EP3}, i.e. for $\nu\to\nu^{EP3}$ and ${\cal E}\to{\cal E}^{EP3}$,
we get $\partial \omega_k/\partial \nu\to \infty$ 
and $\partial \omega_k/\partial {\cal E}\to \infty$, 
leading to $\partial \Gamma/\partial \nu\to \infty$ and $\partial \Gamma/\partial {\cal E}\to \infty$.

%\section{The structure of exceptional points on $\mathrm{\bf{^{85}Rb}}$ and $\mathrm{\bf{^{87}Rb}}$}
%\label{sec:appendix:Rb}
%Under the appropriate conditions, 
%two common isotopes of the Rubidium atom, $\mathrm{^{85}Rb}$ and $\mathrm{^{87}Rb}$, have four optically attainable energy levels 
%\cite{Steck2010Rb85,Steck2010Rb87,Malinovskaya2014Rb}. 
%The upper couple decays with spontaneous emission to the lower couple. 
%These decaying four-level system can be described with the L-GKS equation.

%We located the exceptional points in this system, and got a rich map of \emph{EP}-curves.
%Four of these curves are of particular interest:
%Close to each of the resonances between the two upper and the lower levels, 
%there is an \emph{EP}-curve which is similar to the \emph{EP}-curve we got for the Bloch system.
%These curves can be used for estimating the system parameters: 
%The energy differences, the dipole moment, and the decaying rate.
%Figure \ref{fig:Rb_ep} shows the map of theses four \emph{EP}-curves of $\mathrm{^{87}Rb}$.

\subsection*{Bibliography}
\bibliographystyle{unsrt}
\bibliography{EP_in_Bloch}

\begin{thebibliography}{10}

\bibitem{bloch1946nuclear}
Felix Bloch.
\newblock Nuclear induction.
\newblock {\em Physical review}, 70(7-8):460, 1946.

\bibitem{agarwal1970master}
GS~Agarwal.
\newblock Master-equation approach to spontaneous emission.
\newblock {\em Physical Review A}, 2(5):2038, 1970.

\bibitem{cohen2008optical}
Claude Cohen-Tannoudji, Jacques Dupont-Roc, and Gilbert Grynberg.
\newblock {\em Atom-Photon Interactions: Basic Process and Appilcations},
  chapter Optical Bloch Equations, pages 353--405.
\newblock Wiley-VCH Verlag GmbH, Weinheim, Germany, 1998.

\bibitem{Brewer1983bloch}
Ralph~G. DeVoe and Richard~G. Brewer.
\newblock Experimental test of the optical bloch equations for solids.
\newblock {\em Phys. Rev. Lett.}, 50:1269--1272, Apr 1983.

\bibitem{lloyd1995almost}
Seth Lloyd.
\newblock Almost any quantum logic gate is universal.
\newblock {\em Physical Review Letters}, 75(2):346, 1995.

\bibitem{zrenner2002coherent}
A~Zrenner, E~Beham, S~Stufler, F~Findeis, M~Bichler, and G~Abstreiter.
\newblock Coherent properties of a two-level system based on a quantum-dot
  photodiode.
\newblock {\em Nature}, 418(6898):612--614, 2002.

\bibitem{gammelmark2014fisher}
S{\o}ren Gammelmark and Klaus M{\o}lmer.
\newblock Fisher information and the quantum cram{\'e}r-rao sensitivity limit
  of continuous measurements.
\newblock {\em Physical Review Letters}, 112(17):170401, 2014.

\bibitem{clarke2008superconducting}
John Clarke and Frank~K Wilhelm.
\newblock Superconducting quantum bits.
\newblock {\em Nature}, 453(7198):1031--1042, 2008.

\bibitem{ladd2010quantum}
Thaddeus~D Ladd, Fedor Jelezko, Raymond Laflamme, Yasunobu Nakamura,
  Christopher Monroe, and Jeremy~L OBrien.
\newblock Quantum computers.
\newblock {\em Nature}, 464(7285):45--53, 2010.

\bibitem{lisi2000probing}
E~Lisi, A~Marrone, and D~Montanino.
\newblock Probing possible decoherence effects in atmospheric neutrino
  oscillations.
\newblock {\em Physical Review Letters}, 85(6):1166, 2000.

\bibitem{six82}
J.~Six.
\newblock Test of the non separability of the $k^0\bar{K}^0$ system.
\newblock {\em Physics Letters B}, 114(2):200--202, 1982.

\bibitem{Selleri83}
F.~Selleri.
\newblock Einstein locality and the $k^0\bar{K}^0$ 946-1 0 system.
\newblock {\em Lett. Neovo Cim.}, 36:521, 1983.

\bibitem{privitera1992quantum}
Paolo Privitera and Franco Selleri.
\newblock Quantum mechanics versus local realism for neutral kaon pairs.
\newblock {\em Physics Letters B}, 296(1):261--272, 1992.

\bibitem{datta1986quantum}
Amitava Datta and Dipankar Home.
\newblock Quantum non-separability versus local realism: A new test using the
  b0b0 system.
\newblock {\em Physics Letters A}, 119(1):3--6, 1986.

\bibitem{PhysRevB.90.054304}
R.~Lo~Franco, A.~D'Arrigo, G.~Falci, G.~Compagno, and E.~Paladino.
\newblock Preserving entanglement and nonlocality in solid-state qubits by
  dynamical decoupling.
\newblock {\em Phys. Rev. B}, 90:054304, Aug 2014.

\bibitem{bell2004speakable}
John~S Bell et~al.
\newblock Speakable and unspeakable in quantum mechanics.
\newblock {\em Speakable and Unspeakable in Quantum Mechanics, by JS Bell,
  Introduction by Alain Aspect, Cambridge, UK: Cambridge University Press,
  2004}, 1, 2004.

\bibitem{essen1955atomic}
L~Essen and JVL Parry.
\newblock An atomic standard of frequency and time interval: A c{\ae}sium
  resonator.
\newblock {\em Nature}, 176:280--282, 1955.

\bibitem{giovannetti2004quantum}
Vittorio Giovannetti, Seth Lloyd, and Lorenzo Maccone.
\newblock Quantum-enhanced measurements: beating the standard quantum limit.
\newblock {\em Science}, 306(5700):1330--1336, 2004.

\bibitem{cirac1997improve}
S.~F. Huelga, C.~Macchiavello, T.~Pellizzari, A.~K. Ekert, M.~B. Plenio, and
  J.~I. Cirac.
\newblock Improvement of frequency standards with quantum entanglement.
\newblock {\em Phys. Rev. Lett.}, 79:3865--3868, Nov 1997.

\bibitem{retzker2014increase}
G.~Arrad, Y.~Vinkler, D.~Aharonov, and A.~Retzker.
\newblock Increasing sensing resolution with error correction.
\newblock {\em Phys. Rev. Lett.}, 112:150801, Apr 2014.

\bibitem{wangsness1953dynamical}
Roald~K Wangsness and Felix Bloch.
\newblock The dynamical theory of nuclear induction.
\newblock {\em Physical Review}, 89(4):728, 1953.

\bibitem{bloch1957generalized}
Felix Bloch.
\newblock Generalized theory of relaxation.
\newblock {\em Physical Review}, 105(4):1206, 1957.

\bibitem{davies1974markovian}
E~Brian Davies.
\newblock Markovian master equations.
\newblock {\em Communications in mathematical Physics}, 39(2):91--110, 1974.

\bibitem{kraus1983states}
Karl Kraus.
\newblock {\em States, effects and operations}.
\newblock Springer, 1983.

\bibitem{lindblad1976generators}
Goran Lindblad.
\newblock On the generators of quantum dynamical semigroups.
\newblock {\em Communications in Mathematical Physics}, 48(2):119--130, 1976.

\bibitem{gorini1976completely}
Vittorio Gorini, Andrzej Kossakowski, and Ennackal Chandy~George Sudarshan.
\newblock Completely positive dynamical semigroups of n-level systems.
\newblock {\em Journal of Mathematical Physics}, 17(5):821--825, 1976.

\bibitem{breuer2002theory}
Heinz-Peter Breuer and Francesco Petruccione.
\newblock {\em The theory of open quantum systems}.
\newblock Oxford university press, 2002.

\bibitem{mukamel1999principles}
Shaul Mukamel.
\newblock {\em Principles of nonlinear optical spectroscopy}.
\newblock Number~6. Oxford University Press, 1999.

\bibitem{moiseyev2011non}
Nimrod Moiseyev.
\newblock {\em Non-Hermitian quantum mechanics}.
\newblock Cambridge University Press Cambridge, 2011.

\bibitem{alicki2007qds}
Robert Alicki and Karl Lendi.
\newblock {\em Quantum Dynamical Semigroups and Applications}, volume 717 of
  {\em Lecture Notes in Physics}.
\newblock Springer Berlin Heidelberg, 2007.

\bibitem{fano1961effects}
U.~Fano.
\newblock Effects of configuration interaction on intensities and phase shifts.
\newblock {\em Phys. Rev.}, 124:1866--1878, Dec 1961.

\bibitem{rotter2009nonHermitian}
Ingrid Rotter.
\newblock A non-hermitian hamilton operator and the physics of open quantum
  systems.
\newblock {\em Journal of Physics A: Mathematical and Theoretical},
  42(15):153001, 2009.

\bibitem{berry2004physics}
MV~Berry.
\newblock Physics of nonhermitian degeneracies.
\newblock {\em Czechoslovak journal of physics}, 54(10):1039--1047, 2004.

\bibitem{uzdin2012scattering}
Raam Uzdin and Nimrod Moiseyev.
\newblock Scattering from a waveguide by cycling a non-hermitian degeneracy.
\newblock {\em Physical Review A}, 85(3):031804, 2012.

\bibitem{latinne1995laser}
O~Latinne, NJ~Kylstra, M~D{\"o}rr, J~Purvis, M~Terao-Dunseath, CJ~Joachain,
  PG~Burke, and CJ~Noble.
\newblock Laser-induced degeneracies involving autoionizing states in complex
  atoms.
\newblock {\em Physical review letters}, 74(1):46, 1995.

\bibitem{cartarius2007exceptional}
Holger Cartarius, J{\"o}rg Main, and G{\"u}nter Wunner.
\newblock Exceptional points in atomic spectra.
\newblock {\em Physical review letters}, 99(17):173003, 2007.

\bibitem{uzdin2013effects}
Raam Uzdin, Emanuele~G Dalla~Torre, Ronnie Kosloff, and Nimrod Moiseyev.
\newblock Effects of an exceptional point on the dynamics of a single particle
  in a time-dependent harmonic trap.
\newblock {\em Physical Review A}, 88(2):022505, 2013.

\bibitem{moiseyev2013sudden}
Nimrod Moiseyev.
\newblock Sudden transition from a stable to an unstable harmonic trap as the
  adiabatic potential parameter is varied in a time-periodic harmonic trap.
\newblock {\em Physical Review A}, 88(3):034502, 2013.

\bibitem{rotter1999resonance}
A~I Magunov, I~Rotter, and S~I Strakhova.
\newblock Laser-induced resonance trapping in atoms.
\newblock {\em Journal of Physics B: Atomic, Molecular and Optical Physics},
  32(7):1669, 1999.

\bibitem{rotter2001continuum}
A~I Magunov, I~Rotter, and S~I Strakhova.
\newblock Laser-induced continuum structures and double poles of the s -matrix.
\newblock {\em Journal of Physics B: Atomic, Molecular and Optical Physics},
  34(1):29, 2001.

\bibitem{narevicius2000trapping}
Edvardas Narevicius and Nimrod Moiseyev.
\newblock Trapping of an electron due to molecular vibrations.
\newblock {\em Physical review letters}, 84(8):1681, 2000.

\bibitem{rubinstein2007bifurcation}
J~Rubinstein, P~Sternberg, and Q~Ma.
\newblock Bifurcation diagram and pattern formation of phase slip centers in
  superconducting wires driven with electric currents.
\newblock {\em Physical review letters}, 99(16):167003, 2007.

\bibitem{cejnar2007coulomb}
Pavel Cejnar, Stefan Heinze, and Michal Macek.
\newblock Coulomb analogy for non-hermitian degeneracies near quantum phase
  transitions.
\newblock {\em Physical review letters}, 99(10):100601, 2007.

\bibitem{dembowski2001experimental}
C~Dembowski, H-D Gr{\"a}f, HL~Harney, A~Heine, WD~Heiss, H~Rehfeld, and
  A~Richter.
\newblock Experimental observation of the topological structure of exceptional
  points.
\newblock {\em Physical review letters}, 86(5):787, 2001.

\bibitem{klaiman2008visualization}
Shachar Klaiman, Uwe G{\"u}nther, and Nimrod Moiseyev.
\newblock Visualization of branch points in p t-symmetric waveguides.
\newblock {\em Physical review letters}, 101(8):080402, 2008.

\bibitem{muller2008EP}
Markus M\"uller and Ingrid Rotter.
\newblock Exceptional points in open quantum systems.
\newblock {\em Journal of Physics A: Mathematical and Theoretical},
  41(24):244018, 2008.

\bibitem{muller2009phaseLapses}
Markus M\"uller and Ingrid Rotter.
\newblock Phase lapses in open quantum systems and the non-hermitian hamilton
  operator.
\newblock {\em Phys. Rev. A}, 80:042705, Oct 2009.

\bibitem{Wiersig2014enhancingSensitivity}
Jan Wiersig.
\newblock Enhancing the sensitivity of frequency and energy splitting detection
  by using exceptional points: Application to microcavity sensors for
  single-particle detection.
\newblock {\em Phys. Rev. Lett.}, 112:203901, May 2014.

\bibitem{neuhauser1995FDM}
Michael~R. Wall and Daniel Neuhauser.
\newblock Extraction, through filter-diagonalization, of general quantum
  eigenvalues or classical normal mode frequencies from a small number of
  residues or a short-time segment of a signal. i. theory and application to a
  quantum-dynamics model.
\newblock {\em The Journal of Chemical Physics}, 102(20):8011--8022, 1995.

\bibitem{mandelshtam2001FDM}
Vladimir~A Mandelshtam.
\newblock Fdm: the filter diagonalization method for data processing in nmr
  experiments.
\newblock {\em Progress in Nuclear Magnetic Resonance Spectroscopy},
  38(2):159--196, 2001.

\bibitem{main2000threeHI}
D\v{z}. Belki\'{c}, P.~A. Dando, J.~Main, and H.~S. Taylor.
\newblock Three novel high-resolution nonlinear methods for fast signal
  processing.
\newblock {\em The Journal of Chemical Physics}, 113(16):6542--6556, 2000.

\bibitem{fuchs2014harmonic}
Jacob Fuchs, J{\"o}rg Main, Holger Cartarius, and G{\"u}nter Wunner.
\newblock Harmonic inversion analysis of exceptional points in resonance
  spectra.
\newblock {\em Journal of Physics A: Mathematical and Theoretical},
  47(12):125304, 2014.

\bibitem{noh2010bloch}
Heung-Ryoul Noh and Wonho Jhe.
\newblock Analytic solutions of the optical bloch equations.
\newblock {\em Optics Communications}, 283(11):2353 -- 2355, 2010.

\bibitem{moroz2012unorthodox}
Alexander Moroz.
\newblock On unorthodox solutions of the bloch equations.
\newblock {\em arXiv preprint arXiv:1208.5736}, 2012.

\bibitem{noh2015bloch}
Heung-Ryoul Noh.
\newblock Optical bloch equations for a two-level atom revisited: Analytical
  solutions.
\newblock {\em Journal of the Physical Society of Japan}, 84(9):094402, 2015.

\bibitem{berry1979elliptic}
Michael~V Berry, JF~Nye, and FJ~Wright.
\newblock The elliptic umbilic diffraction catastrophe.
\newblock {\em Philosophical Transactions for the Royal Society of London.
  Series A, Mathematical and Physical Sciences}, pages 453--484, 1979.

\bibitem{Mailybaev2006computation}
Alexei~A. Mailybaev.
\newblock Computation of multiple eigenvalues and generalized eigenvectors for
  matrices dependent on parameters.
\newblock {\em Numerical Linear Algebra with Applications}, 13(5):419--436,
  2006.

\bibitem{geva1995relaxation}
Eitan Geva, Ronnie Kosloff, and JL~Skinner.
\newblock On the relaxation of a two-level system driven by a strong
  electromagnetic field.
\newblock {\em The Journal of chemical physics}, 102(21):8541--8561, 1995.

\bibitem{szczygielski2013markovian}
Krzysztof Szczygielski, David Gelbwaser-Klimovsky, and Robert Alicki.
\newblock Markovian master equation and thermodynamics of a two-level system in
  a strong laser field.
\newblock {\em Physical Review E}, 87(1):012120, 2013.

\bibitem{Benko2008noisyFDM}
Uro{\v{s}} Benko and {\DJ}ani Juri{\v{c}}i{\'c}.
\newblock Frequency analysis of noisy short-time stationary signals using
  filter-diagonalization.
\newblock {\em Signal Processing}, 88(7):1733 -- 1746, 2008.

\bibitem{hu199876FDM_noise}
Haitao Hu, Que~N. Van, Vladimir~A. Mandelshtam, and A.J. Shaka.
\newblock Reference deconvolution, phase correction, and line listing of {NMR}
  spectra by the 1d filter diagonalization method.
\newblock {\em Journal of Magnetic Resonance}, 134(1):76 -- 87, 1998.

\bibitem{Celik2010FDM_sensitivity}
Hasan Celik, A.J. Shaka, and V.A. Mandelshtam.
\newblock Sensitivity analysis of solutions of the harmonic inversion problem:
  Are all data points created equal?
\newblock {\em Journal of Magnetic Resonance}, 206(1):120 -- 126, 2010.

\bibitem{martini2014mass_spectrometry}
Beau~R. Martini, Konstantin Aizikov, and Vladimir~A. Mandelshtam.
\newblock The filter diagonalization method and its assessment for {F}ourier
  transform mass spectrometry.
\newblock {\em International Journal of Mass Spectrometry}, 373(0):1 -- 14,
  2014.

\bibitem{morag2015EPatomic}
Morag Am-Shallem, Ronnie Kosloff, and Nimrod Moiseyev.
\newblock Parameter estimation in atomic spectroscopy using exceptional points.
\newblock {\em arXiv preprint arXiv:1511.07205}, 2015.

\bibitem{emch1979standard}
G{\'e}rard~G Emch and Joseph~C Varilly.
\newblock On the standard form of the bloch equation.
\newblock {\em Letters in Mathematical Physics}, 3(2):113--116, 1979.

\bibitem{fogli2007probing}
GL~Fogli, E~Lisi, A~Marrone, D~Montanino, and A~Palazzo.
\newblock Probing nonstandard decoherence effects with solar and kamland
  neutrinos.
\newblock {\em Physical Review D}, 76(3):033006, 2007.

\bibitem{gammelmark2014hidden}
S~Gammelmark, K~M{\o}lmer, W~Alt, T~Kampschulte, and D~Meschede.
\newblock Hidden markov model of atomic quantum jump dynamics in an optically
  probed cavity.
\newblock {\em Physical Review A}, 89(4):043839, 2014.

\end{thebibliography}

\end{document}